\documentclass[10pt]{article}
\usepackage{amsthm, amsmath , amssymb, eucal, color} 
\usepackage{graphics,float}

\makeatletter
\renewcommand{\@oddhead}{\it 
   Fass\`o and Sansonetto: Ball in a rotating surface and kasamawashi
\hfill \thepage}
\makeatother

\newtheorem{theorem}{\bf Theorem}
\newtheorem{proposition}[theorem]{\bf Proposition}

\newtheorem*{remark}{\bf Remark}

\makeatletter
\renewcommand\subsection{\@startsection{subsection}{0}{\z@}%
                                     {-3.25ex\@plus -1ex
\@minus-.2ex}%
                                     {0ex}%
                                     {\normalfont\bfseries}}
\makeatother

\newcommand\PS[2]{\,#1\!\cdot\!#2\,}

\newcommand\bList{
\begin{list}{}{\leftmargin2em\labelwidth1.5em\labelsep.5em\itemindent0em
\topsep.3ex\itemsep-.4ex} }
\newcommand\eList{\end{list}}

\newcommand{\cE}{\mathcal{E}}

\newcommand{\cL}{\mathcal{L}}

\newcommand{\cR}{\mathcal{R}}

\newcommand{\reali}{\mathbb{R}}
\newcommand{\bR}[1]{\mathbb{R}^{#1}}

\renewcommand{\O}{\Omega}

\newcommand{\SO}[1]{\textrm{SO{(#1)}}}

\newcommand\x{\mathrm{x}}
\newcommand\y{\mathrm{y}}
\newcommand\z{\mathrm{z}}
\newcommand\X{\mathrm{X}}
\newcommand\Y{\mathrm{Y}}
\newcommand\Z{\mathrm{Z}}

\renewcommand{\a}{\alpha}

\renewcommand{\o}{\omega}

\renewcommand\S{\Sigma}

\title{\bf On some aspects of the dynamics \\
of a ball in a rotating surface of revolution\\
and of the kasamawashi art}

\author{Francesco Fass\`o\footnote{
Universit\`a degli Studi di Padova,
Dipartimento di Matematica Tullio Levi-Civita,
via Trieste 63, 35121 Padova.
{\tt (E-mail: fasso@math.unipd.it)}.
} 
\and
Nicola Sansonetto \footnote{
Universit\`a degli Studi di Verona, 
Dipartimento di Informatica, 
Strada le Grazie 15, 37134 Verona, Italy.
{\tt (E-mail: nicola.sansonetto@univr.it)}.
}
}

\date{\small (\today)}
\begin{document}
\maketitle

\vskip-2mm
\hspace{7.5cm}{\it Dedicated to memory of Alexey V. Borisov}
\vspace{.4cm}

{\small
\begin{abstract}
\noindent

We study some aspects of the dynamics of the nonholonomic system
formed by a heavy homogeneous ball constrained to roll without sliding
on a steadily rotating surface of revolution. First, in the case in
which the figure axis of the surface is vertical (and hence the
system is $\textrm{SO(3)}\times\textrm{SO(2)}$-symmetric) and the surface has a
(nondegenerate) maximum at its vertex, we show the existence of
motions asymptotic to the vertex and rule out the possibility of blow
up. This is done passing to the 5-dimensional $\textrm{SO(3)}$-reduced system.
The $\textrm{SO(3)}$-symmetry persists when the figure axis of the surface is
inclined with respect to the vertical---and the system can be viewed
as a simple model for the Japanese kasamawashi (turning umbrella)
performance art---and in that case we study the (stability of the)
equilibria of the 5-dimensional reduced system.

\vskip 1truecm
\end{abstract}
}

{\small


}

\section{Introduction}

The system
formed by a heavy homogeneous ball that rolls without sliding on a
surface of revolution, which either is at rest or steadily rotates
around its vertical figure axis with constant angular velocity $\O$,
is a classical system of nonholonomic mechanics. Its first studies go
back to Routh, and there has recently been a renew of interest, see
e.g. \cite{hermans,zenkov,BMK2002,FGS,FG2007,
FS2016,BMB2015,BIKM2018,BIKM2019,BIKM2020,BalsSans,BY,dVFS}. A rather general
study of the dynamics of the system has been the object of the very
recent article \cite{dVFS}, which is the basis for the present study.

The system has an 8-dimensional phase space and an
$\SO3\times\SO2$-symmetry (rotate the ball about its center and the
center about the surface's figure axis). Reduction can be done in
stages, obtaining first a $5$-dimensional $\SO3$-reduced system and
then a $4$-dimensional $\SO3\times \SO2$-reduced system. Most of the
above analyses have been performed in either reduced system. The
5-dimensional reduced system looses the information on the attitude
of the ball and describes the motion of the center (or of the
contact point) of the ball along the surface and of the ball's angular
velocity. Specifically, a possible choice of the five coordinates in
the $\SO3$-reduced space are the horizontal coordinates and
velocities of the center of the ball and the vertical component of the
angular velocity vector (the other two components of the angular
velocity vector are then determined by the rolling
constraint). The 4-dimensional reduced system neglects also
the rotation of the center of the ball around the surface's figure
axis and describes only the radial motion of the center of the ball
and, again, the angular velocity. 

The unreduced system has three independent $\SO3\times\SO2$-invariant
first integrals, which are inherited by both reduced systems. One is 
the energy if $\O=0$ and a generalization of it called `moving energy'
if $\O\not=0$ \cite{FS2016,dVFS}. The existence of the other two was
proven by Routh if $\O=0$ \cite{routh} and by Borisov, Kilin and
Mamaev \cite{BMK2002} if $\O\not=0$. Therefore, the 4-dimensional
$\SO3\times\SO2$-reduced system has three independent integrals of
motion, and this has made it possible to prove a number of results on
its dynamics. In particular, if the surface on which the ball rolls
goes to $+\infty$ at infinity (sufficiently fast, if $\O\not=0$),
then the common level sets of these three integrals in the
4-dimensional reduced phase space are compact and the dynamics of the
4-dimensional reduced system is generically periodic;
correspondingly, reconstruction results for relative periodic orbits
of symmetric systems with compact symmetry groups (which date back to
the 1980's and are due to Krupa and Field \cite{field, krupa}, see
also \cite{hermans,FG2007,CDS}) ensure that the dynamics of the 5-dimensional
reduced system is generically almost-periodic on tori of dimension 2
and that of the unreduced system is generically almost-periodic on
tori of dimension 3. This result was proven in the 1990s by Hermans
\cite{hermans} and Zenkov \cite{zenkov} in the case $\O=0$, but its
extension to the case of a rotating surface \cite{FS2016,dVFS} had to
wait for the discovery of the conservation of the moving energy
because the energy is (except in special situations \cite{FS2015})
not conserved for a nonholonomic system with nonhomogeneous
constraints. 

The study of the 4-dimensional reduced system benefits of the fact
that, thanks to the existence of a rank-two Poisson structure that
makes the system Hamiltonian (\cite{BMK2002} for $\O=0$, \cite{dVFS} for
$\O\not=0$) its phase space is foliated by two-dimensional invariant
submanifolds on which the dynamics is Hamiltonian (and even
Lagrangian). This allowed to study and classify, for instance, its
equilibria \cite{dVFS}. Numerical investigations of the reduced dynamics
in the particular case of a rotating conical surface are given in
\cite{BIKM2020}.

\subsection{The dynamics near the vertex. }
Even though very successful, the analysis in the 4-dimensional
$\SO3\times\SO2$-reduced space has a limitation due to the fact that
the $\SO2$-action is not free (the rotation about the figure axis
keeps fixed all kinematical states in which the center of the ball is
at the vertex of the surface with zero velocity---and the ball has
any vertical angular velocity) and the $\SO2$-reduced space is
singular. This complicates the study of motions
in which the ball passes through the vertex, which to our knowledge
has never been undertaken so far. 

Of course, it is intuitively clear that, whichever the geometry of
the surface\footnote{As long as it is regular at the vertex, thus
excluding e.g. the case of a conical surface} and its rotational
velocity $\O$, the $4$-dimensional reduced system has equilibria that
correspond to the ball sitting at the vertex and spinning with any
vertical angular velocity. However, their stability has not been
investigated so far. In particular, it is not known if there are
motions asymptotic to such equilibria at the vertex. Reference
\cite{dVFS} points out that, particularly when $\O\not=0$, it is not
even ruled out the possibility of `blow up' at the vertex, namely, of
motions in which the center of the ball approaches (or even reaches
in finite time) the vertex with the angular velocity of the ball that
goes to infinity. 

The main objective of the present article is to give some answers to
these questions. Following an indication in \cite{dVFS}, we will do
it by studying the $5$-dimensional $\SO3$-reduced system, whose phase
space is regular at the vertex. We will first of all prove that there
is no possibility of blow up at the vertex. Next, we will investigate
the reduced equilibria of the 5-dimensional reduced system that
correspond to the ball sitting at the vertex. Quite clearly, there is
a one-parameter family of them (parametrized by the vertical component
of the ball's angular velocity) and this implies that their Lyapunov
stability may be elusive. Nevertheless, the study of the
linearization at these equilibria gives important information,
because the presence of eigenvalues with negative (positive) real
part implies the existence of a stable (unstable) manifold and hence
of motions asymptotic to the vertex for $t\to+\infty$
($t\to-\infty$). We will show that, if the surface has a (local or
global) nondegenerate maximum at the vertex, then motions of this type do exist.
In addition, we will study some aspects of the stability of the 
reduced equilibria at the vertex. 

\subsection{Kasamawashi, or the ball on a rotating umbrella. }
We take the opportunity of approaching this study from a more general
perspective and consider the more general case in which the figure
axis of the surface of revolution on which the ball rolls may also
be inclined of a certain angle $\a$ with respect to the vertical. For
$\a=0$ we have the system described above. The system with $\a\not=0$
does not appear to have been investigated so far, except in the case in
which the surface is a plane \cite{BIKM2018}. 

If $\a\not=0$ the system looses the $\SO2$-symmetry (except for
special geometries of the surface, such as that of a sphere)
but retains its $\SO3$-symmetry. It is
therefore possible to consider the $5$-dimensional $\SO3$-reduced
system. We do not undertake here a systematic study of the dynamics
of this reduced system, which if $\a\not=0$ can be expected to be
nonintegrable. However, as a slight extension of our study of the
case $\a=0$ we will investigate the equilibria of the $\SO3$-reduced
system and their stability. We shall show that the only equilibria of
such reduced system correspond to motion of the unreduced system in
which the center of the ball stays fixed in space, touching the
surface at a point at which the tangent plane is horizontal
(due to the rotational symmetry of the surface, the contact takes
place at a point that changes in the surface but stays fixed in
space), and spins with any vertical angular velocity. We shall
analyze the spectral stability of these reduced equilibria.

It is tempting, if not even natural, to relate this analysis to the
Japanese {\it kasamawashi} (``turning umbrella'') art, in which a ball
(or a disk or ring) is posed on a tilted conic umbrella, that the performer rotates
so as to keep the ball at the same spatial position. The art is very
fascinating and its modelling, of course, is a matter of control (the
realization of a robot that performs kasamawashi through a PID
controller has been reported in \cite{WKK}, without however any
mathematical or modelling detail). Nevertheless, this purely dynamical
approach seems capable of giving some information.

\section{The system}

\subsection{The nonholonomic system}

We follow the description of the system given in \cite{dVFS}, which
however considers only the case $\a=0$ (and, less important, the case
in which the surface is a graph over $\bR2$, namely $D=\bR2$
in the notation below). We begin considering the
holonomic system formed by a homogeneous ball of mass $m$ and radius
$a$ whose center $C$ is constrained to a smooth surface of revolution
$\Sigma$ which is embedded in $\bR{3}\ni(\X,\Y,\Z)$ with its vertex
at the origin and its figure axis that forms an angle $\alpha$,
$0\le\alpha<\frac\pi2$, with the $\Z$-axis, which is the ascending
vertical. 
We describe the system with respect to a (spatial) reference frame
$\{O;\x,\y,\z\}$ with the origin $O$ at the vertex of $\S$, the $\z$-axis
aligned with the figure axis of $\S$, the $\y$-axis horizontal and
the $\x$-axis tangent to $\S$ at the vertex, see Fig. 1. We
parametrize the system with the rescaled coordinates
$x=(x_1,x_2)=(\frac \x a,\frac \y a)$ and describe the surface $\S$,
in the frame $\{O;\x,\y,\z\}$, via the parametrization
$$
  D \ni x=(x_1,x_2) \mapsto \big( a x_1, a x_2 , a f(|x|) \big) 
$$
where $D=\{x\in\bR2\,:\; |x|<R\}$ for some $R>0$ or $R=+\infty$ and
$f:I\to\reali$ with $I=(-R,R)$ is an even smooth function that we call the
{profile} function ($|\ |$ denotes the Euclidean norm in $\bR2$).
Obviously, $f'(0)=0$.


\begin{figure}[h]
\begin{center}
{\small
{\scalebox{.5}{\includegraphics*{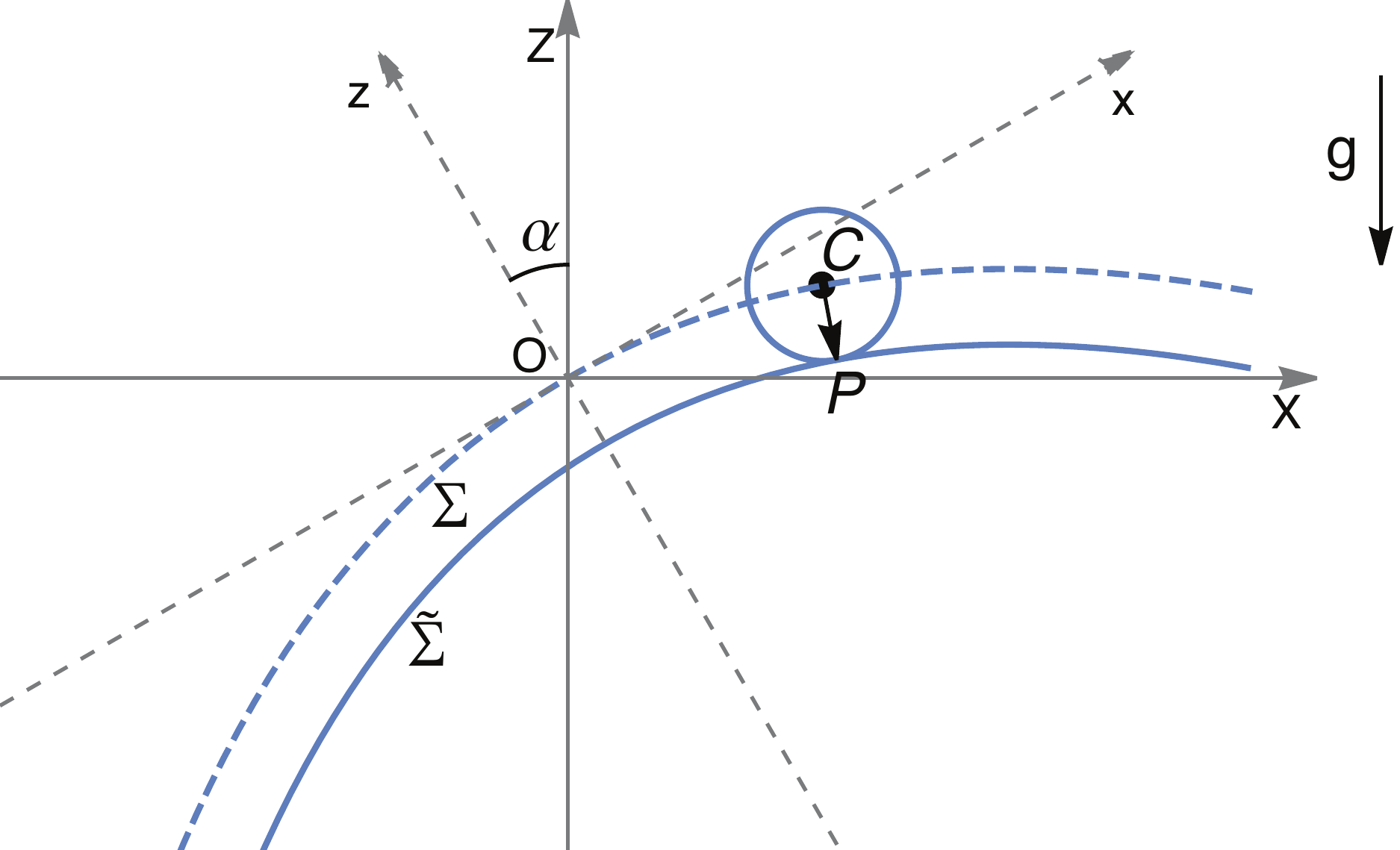}}}
}
\caption{\small The generatrices $\Gamma$ and $\tilde\Gamma$ of the
surfaces $\Sigma$ and $\tilde\Sigma$.}
\end{center}
\end{figure}



Since smoothness at $x=0$ of the function $x\mapsto f(|x|)$ is not
manifest, and we are specifically interested to the dynamics near
that set, following \cite{FGS,dVFS} we will use instead a smooth
function $\psi:\reali\to\reali$ such that
$$
  f(r) = \psi\big( {\textstyle\frac12} r^2 \big)
  \qquad
  \forall r\in I \,.
$$
The existence of such a function is granted by a result of Whitney
\cite{whitney} (see also \cite{GG}, pages 103 and 108, and
\cite{FGS}) on account of the fact that $f$ is even. Note that
\begin{equation}
\begin{aligned}
\label{psi'psi''}
  &\psi'\big({\textstyle\frac12}r^2\big) =
  \frac{f'(r)}r
  \qquad \mathrm{for\ } r>0
  \,,\qquad
  &&\psi'(0) = f''(0) \,,
  \\
  &\psi''\big({\textstyle\frac12}r^2\big)=\frac{rf''(r)-f'(r)}{r^3}
  \qquad \mathrm{for\ } r>0
  \,,\qquad
  &&\psi''(0) = \frac13f^\mathrm{(iv)}(0) \,.
\end{aligned}
\end{equation}
In stating our results, we will however use as much as possible the
profile function $f$ and its derivatives, whose interpretation is more
direct.

The configuration manifold of this holonomic system can be identified
with $D\times \SO3\ni(x,\cR)$, where $\cR$ is the matrix that turns the
spatial frame $\{O;\x,\y,\z\}$ into some chosen body frame, and,
after (right) trivialization of the tangent bundle
of $SO(3)$, its phase space can be identified with the 10-dimensional manifold
\[
  M_{10}
  =
  D \times \SO{3}\times \bR{2}\times\bR{3} \ni (x,\cR, v,\o)\,
\]
where $v=(v_1,v_2)=\dot x$ and $\o=(\o_\x,\o_\y,\o_\z)$ is the angular velocity of the ball
relative to, and written in, the frame $\{O;\x,\y,\z\}$. 

We assume that the only active force that acts on the system is
weight, directed as the downward vertical $Z$-axis, and denote by $a
\hat g$ the gravity acceleration and by $mka^2$, with some $0<k<1$,
the moment of inertia of the ball with respect to its center.
Thus, since $Z=\x\sin\a+a\psi\big({\textstyle\frac12}|x|^2\big)\cos\a$
in the points of $\S$ and the velocity $V_C$ of $C$ equals
$a\big(v_1,v_2, \PS xv \psi'({\textstyle\frac12}|x|^2)\big)$, the
Lagrangian of the holonomic system is
\begin{equation}
\label{eq:Lagrangian}
  \cL(x,\cR,v,\o) =   \frac12|v|^2
  +
  \frac12 \big( \PS xv \psi'\big({\textstyle\frac12} |x|^2\big) \big)^2
  +
  \frac12 k |\o|^2
  -
  \hat g \big( x_1\sin\a
           + \psi\big({\textstyle\frac12} |x|^2\big) \cos\a \big) \,.
\end{equation}

We now add to the system the nonholonomic constraint that the ball
rolls without sliding on a surface $\widetilde\Sigma$ which lies {\it
below} $\S$, is parallel to it, and rotates with constant angular
velocity $\Omega e_\z$ about its figure axis, namely, the $\z$-axis.
We assume that
\begin{equation}
\label{eq:curvature}
    f''(r) > - (1+f'(r)^2)^{3/2} \qquad \forall r \in I \,,
\end{equation}
which ensures that $\tilde \S$ is regular and that, in
any configuration, the ball touches $\tilde\S$ in a single point, see
\cite{dVFS}. 

Since the point $P$ of the ball in contact with $\tilde\S$ has
velocity $V_C + \o \times CP$ and the point of $\tilde \S$ with
which $P$ is in contact has velocity $\O e_z\times OP$, the
nonholonomic constraint is given by 
\begin{equation}
\label{vinc1}
  V_C + \o \times CP = \Omega e_\z \times OP \,.
\end{equation}
Here, $OP=OC+CP = (ax_1,ax_2,a \psi\big({\textstyle\frac12}|x|^2\big))
 +a n(x)$ with $n(x)$ the
(downward) normal unit vector to $\Sigma$ at its point $C$, namely
$
n(x) = \frac1{F(|x|)}
\big(
x_1 \psi'({\textstyle\frac12}|x|^2 ),
x_2 \psi'({\textstyle\frac12}|x|^2 ),
-1
\big)
$ 
with the function $F:I\to\reali$ defined as
\begin{equation}
  F(r) 
  \;:=\; \sqrt{ 1 +f'(r)^2} 
  = \sqrt{ 1 +r^2\psi'({\textstyle\frac12}r^2)^2 } 
  \,.
\end{equation}
Thus, the first two entries of \eqref{vinc1} can be
written\footnote{From now it is understood that, unless differently
specified, $\psi$ and its
derivatives are evaluated at ${\textstyle\frac12}|x|^2$ and $F$ at
$|x|$.}
\begin{equation}
\label{vinc2}
  \o_\x  = \; 
  -Fv_2   -x_1 \psi'\o_\z +\O x_1(1+\psi')
  \,, \quad
  \o_\y = \;
  Fv_1 -x_2 \psi' \o_\z  +
  \O x_2(1 +\psi') 
\end{equation}
(the third entry is not independent) and define an $8$-dimensional
submanifold of $M_{10}$ which is diffeomorphic to
$$
  M_8 = D \times\SO{3}\times\bR{2}\times\bR{} 
        \ni (x,\cR,\dot x,\o_\z) 
$$
and is the phase space of the nonholonomic system.  

The equations of motion of the nonholonomic system in $M_8$ can be
obtained with various standard techniques, and are the five equations
\begin{equation}
\label{eq:eqmoto}
\begin{aligned} 
  \dot x_1 = & v_1
  \\
  \dot x_2 = & v_2
  \\  
  \dot v_1 = & 
    - \frac1{F^2}
    \Big[
      \gamma\big(x_1\psi'\cos\a+(1+x_2^2\psi'^2)\sin\a\big)
      -\mu\big(v_2\psi'+x_2 \PS xv \psi''\big)\o_\z F
      \\
      &\qquad\quad
      +\mu v_1 \PS xv\big(\psi'+|x|^2 \psi''\big)\psi'
      +\frac{x_1}{1\!+\!k}
         \big(|v|^2\psi'+(x\!\cdot\!v)^2\psi''\big)\psi'
    \\
    &\qquad\quad
    -\O \mu 
    \big[
       v_2F(F+\psi')
       +x_2 \PS xv\big(\psi'^2+|x|^2\psi'\psi'' +F\psi''\big) 
    \big]
    \Big]
  \\  
  \dot v_2 = &  
    -\frac1{F^2}
    \Big[
      \gamma x_2 \psi'(\cos\a-x_1\psi'\sin\a)
      +\mu\big(v_1\psi'+x_1 \PS xv\psi''\big)\o_\z F 
      \\
      &\qquad\quad
      +\mu v_2 \PS xv \big(\psi'+|x|^2 \psi''\big)\psi'
      +\frac{x_2}{1\!+\!k}\big(|v|^2\psi'+(\PS xv)^2\psi''\big)\psi'
    \\
    &\qquad\quad
    + \O \mu 
    \big[
       v_1F(F+\psi')
       +x_1 \PS xv\big(\psi'^2+|x|^2\psi'\psi'' +F\psi''\big) 
    \big]
    \Big] 
  \\
  \dot \o_\z = &
    -\frac1F \gamma x_2\psi'\sin\a
    -\frac{\PS xv \psi'}{(1\!+\!k)F^3} 
    \Big[
    \big(\o_\z F+(x_1v_2-x_2v_1)\psi'\big) \big(\psi'+|x|^2\psi''\big) 
    \\
    &\qquad\quad
    -\O 
    \big[F^2 + (F+|x|^2\psi')\big(\psi'+|x|^2\psi''\big)\big]
    \Big]
\end{aligned}
\end{equation}
where $\gamma=\frac{\hat g}{1+k}$ and $\mu=\frac k{1+k}$, completed
with the restriction to $M_8$ of the equation $\dot \cR=\hat\o\cR$
with $\hat \o$ the antisymmetric matrix associated to
the vector $(\o_\x,\o_\y,\o_\z)\in\bR3$
(with $\o_\x$ and $\o_\y$ as in \eqref{vinc2}). Some
indications on how to obtain these equations are given in the
Appendix.

\begin{remark}
This formulation assumes smoothness of the surface
$\S$. In certain cases---such as that of a cone---the surface is not
smooth at the vertex. In such cases, Eqs. \eqref{eq:eqmoto}
describe the motions outside a neighbourhood of the vertex. Thus,
they can be used to study the equilibria of the system at locations
different from the vertex, which is what we will do for an inclined
conic surface in section \ref{s:kasam}. 
\end{remark}

\subsection{The $\mathbf{\SO3}$-reduced system }
\label{ss:reduction}

Consider now the right action $\Xi$ of $\SO{3}$ on $M_{10}$ on the
$\SO{3}$-factor: $\Xi_{S}(x,\cR,\dot x,\o) = (x,\cR S,\dot
x,\o)$. From \eqref{vinc2} it follows that the constraint
manifold $M_8$ is invariant under the action $\Xi$ and thus $\Xi$
restricts to an action on $M_8$. Since the Lagrangian
\eqref{eq:Lagrangian} as well is invariant under $\Xi$, the equations
of motion of the nonholonomic system in $M_8$ can be reduced to
$M_8/\SO{3}$ \cite{BS,BKMM}. Since the Lagrangian and the constraint
are independent of the attitude $\cR$ of the ball, the
$\SO{3}$-reduction consists in simply cutting off the factor $\SO3$
of $M_8$. Thus, the $\SO{3}$-reduced space is the five-dimensional
manifold
$$
  M_5 = D\times\bR{2}\times\bR{}\ni(x,v,\o_\z) 
$$
and the equations of motion of the reduced system are Eqs.
\eqref{eq:eqmoto}. These equations define a vector field on $M_5$.

Note that the motion $t\mapsto(x(t),v(t),\o_\z(t))$ of the
$\SO3$-reduced system determines the motion
$t\mapsto(x(t),\cR(t),v(t),\o_\z(t))$ of the unreduced system except
for the attitude $t\mapsto \cR(t)$ of the ball, which can in principle
be determined via the ``reconstruction equation'' $\dot
\cR(t)=\hat\o(t) \cR(t)$, where
$t\mapsto\o(t)=(\o_\x(t),\o_\y(t),\o_\z(t))$ with the first two
components determined by the constraint Eq. \eqref{vinc2}.

\section{The equilibria of the $\mathbf{\SO3}$-reduced system}
\label{s:equilibria} 

\subsection{The $\SO3$-reduced equilibria. }
We determine now the equilibria of the $\SO3$-reduced system. 

\begin{proposition}
\label{p:equilibria} 
The equilibria of the $\SO3$-reduced system are the points
$(x,0,\o_\z)\in M_5$ with any $\o_\z\in\bR{}$ and any $x$ such that
the normal to the surface $\S$ at the point of coordinate $x$ has
horizontal tangent plane, namely:
\bList
\item[i.] If $\a=0$, $x$ such that $f'(|x|)=0$. 
\item[ii.] If $\a\not=0$, $x_2=0$ and $x_1$ such that
$f'(|x_1|)=-\mathrm{sign}(x_1)\tan(\a)$. 
\eList
\end{proposition}

\begin{proof}
At an equilibrium, $v_1=v_2=0$ and the vanishing of $\dot v_1$, $\dot v_2$ and
$\dot \o_\z$ in \eqref{eq:eqmoto} gives the three conditions 
\begin{equation}
\label{condizioni}
\begin{aligned} 
  & x_1\psi'\big({\textstyle\frac12}|x|^2\big)\cos\a +
  \big(1+x_2^2\psi'({\textstyle\frac12}|x|^2)^2\big) \sin\a = 0
  \\
  & x_2\psi'\big({\textstyle\frac12}|x|^2\big)
  \big( x_1\psi'\big({\textstyle\frac12}|x|^2\big)\sin\a-\cos\a\big) = 0
  \\
  &x_2 \psi'\big({\textstyle\frac12}|x|^2\big)\sin\a = 0
\end{aligned} 
\end{equation}
on $x=(x_1,x_2)$. Since $\o_\z$ does not enter them, it is arbitrary at
the equilibria. 

If $\a=0$, then the last condition \eqref{condizioni}
is satisfied for all $x$ while
the first two give $x_1\psi'\big({\textstyle\frac12}|x|^2\big)=
x_2\psi'\big({\textstyle\frac12}|x|^2\big)= 0$. These two
conditions are satisfied at all points at which $x=0$ and/or
$\psi'\big({\textstyle\frac12}|x|^2\big)=0$, namely, as follows from
\eqref{psi'psi''}, all points at which $f'(|x|)=0$.

If $\a\not=0$, then the last condition \eqref{condizioni} is
satisfied if $x_2=0$ and/or if $\psi'\big({\textstyle\frac12}|x|^2\big)=0$.
But in the latter case
the first condition \eqref{condizioni} is never satisfied because
$\sin\a\not=0$. If $x_2=0$ then the second condition
\eqref{condizioni} is satisfied by all $x_1$ and the first one
reduces to $x_1\psi'\big({\textstyle\frac12}{x_1^2}\big)\cos\a+\sin\a =0$.
Since $\sin\a\not=0$,
necessarily $x_1\not=0$ and
$f'(|x_1|)\mathrm{sign}(x_1)=x_1\psi'\big({\textstyle\frac12}{x_1^2}\big)
=-\tan\a$.

The normal to $\S$ at the point $C$, in the frame $\{O;X,Y,Z\}$, is 
$\left(\begin{matrix} 
\cos\a & 0 & \sin\a \\ 0 & 1 & 0 \\ -\sin\a & 0 & \cos\a 
\end{matrix}\right)n(x)$
and the vanishing of its first two components is equivalent to
$x_2=0$, $\sin\a+x_1 \psi'\big({\textstyle\frac12}{x_1^2}\big) \cos\a=0$.
\end{proof}

The $\SO3$-reduced equilibria reconstruct to ($\SO3$-families of)
motions of the unreduced system in which the ball `sits' at a point
in space and either spins around its center or stays still.
These families of motions form the so-called relative equilibria 
of the unreduced system. It follows from the already mentioned
reconstruction theory of Krupa and Field that, since $\SO3$ is
compact and has rank one, all motions of the ball in a relative
equilibrium are  periodic (or, as a particular case, equilibria,
which happens if $\o_\z=\O=0$).

Since $f'(0)=0$, when $\a=0$ there is always a family of reduced
equilibria with $x=0$ and any $\o_\z$, that we call ``reduced
equilibria at the vertex''. 

In addition, when $\a=0$, there are families of reduced equilibria
with any $\o_\z\in\reali$ and any $x$ in a `critical parallel' of the
surface $\S$, namely the parallels on which $f'=0$. We note that the
existence of these reduced equilibria was already proven in
\cite{dVFS}. Specifically, the equilibria of ``type RE2'' of the
$\SO3\times\SO2$-reduced system found in \cite{dVFS} reconstruct
exactly to these equilibria of the $\SO3$-reduced system (see
particularly section 5.2 of \cite{dVFS}). Since their (spectral)
stability properties have already been investigated in \cite{dVFS},
we will not consider them here anymore.

When $\a\not=0$, instead, the reduced equilibria reconstruct to
periodic orbits (equilibria) of the unreduced system in which the
ball spins around the vertical (stays still) and touches the surface
at a point at which the tangent plane to the surface is horizontal
and stays fixed in space. Note that, if $\a\not=0$, the contact point
at such reduced equilibrium is never at the vertex of $\tilde\S$.

\begin{remark}
It follows from the reconstruction of the equilibria
of the $\SO3\times\SO2$-reduced system in \cite{dVFS} that, for
$\a=0$, the $\SO3$-reduced system possesses periodic orbits in which
the center of the ball moves steadily on {\it any} parallel of the
surface.
\end{remark}

\subsection{Linearization. }
Since in the $\SO3$-reduced equations of motion \eqref{eq:eqmoto} the
coordinate $\o_\z$ is always multiplied by either $v_1$ or $v_2$, the
last column of the Jacobian matrix of the $\SO3$-reduced vector field
vanishes at the equilibria. Therefore, the linearization at a reduced
equilibrium has always an eigenvalue $0$. Its presence is related to
the fact that the reduced equilibria all come in families,
parametrized by $\o_\z\in\reali$. The remaining four eigenvalues are
determined by the first $4\times 4$ block of the linearization matrix.

As already said, when $\a=0$ we exclude from our consideration
the reduced equilibria with $x\not=0$.\footnote{They form
two-parameter families and therefore there at least two zero
eigenvalues of the linearization. But in fact, there are always three
zero eigenvalues; this can be explained through the already mentioned
fact that the $\SO3\times\SO2$-reduced system has a Hamiltonian
structure.}  In the remaining equilibria $x_2=0$ and the first
$4\times4$ block of the linearization matrix at the equilibrium
$(x_1,0,0,0,\o_\z)$ has the form 
\begin{equation}
\label{eq:linearisation_5}
  \begin{pmatrix}
  0 & 0 & 1 & 0 \\
  0 & 0 & 0 & 1 \\
   a_{31} & 0 & 0 &  a_{34} \\
  0 &  a_{42} &  a_{43} & 0 \\
  \end{pmatrix}
\end{equation}
with 
\begin{equation}
\label{a}
\begin{aligned} 
  &a_{31} = \frac\gamma{F^4} \big(\psi'+x_1^2\psi''\big) 
    \big(2x_1\psi'\sin\a+(x_1^2\psi'^2-1)\cos\a\big)
  \\
  & a_{34} = \frac\mu F  \psi'\o_\z
    -\O \frac\mu {F^2} \big(1+F\psi'+x_1^2\psi'^2\big)
  \\
  & a_{42} = \frac\gamma{F^2}\psi'
    \big(x_1\psi'\sin\a -\cos\a\big) 
  \\
  & a_{43} = -\frac\mu F \big(\psi'+x_1^2\psi''\big)\o_\z
  + \O \frac\mu{F^2}
   \big( F^2 +(x_1^2\psi'+F)\psi'+ x_1^2(F+x_1^2\psi')\psi''\big) \,. 
\end{aligned}
\end{equation}
where $\psi'$ and $\psi''$ are evaluated at
${\textstyle\frac12}x_1^2$ and $F$ at $x_1$. The
characteristic polynomial of this matrix is the biquadratic polynomial
\begin{equation}
\label{cp}
  \lambda^4 - (a_{31}+a_{42}+a_{34}a_{43})\lambda^2 + a_{31}a_{42} \,.
\end{equation}

\section{The dynamics near the vertex in the case $\mathbf{\a=0}$ } 

In this section we consider the system formed by the ball
nonholonomically constrained to the surface with $\a=0$. The main
question is whether there exist motions in which, asymptotically, the
ball tends to the vertex.

\subsection{No blow up at the vertex. }
First, we show that no such motions exists in which the angular
velocity $\o_\z$ explodes. This answers a question raised in
\cite{dVFS}. This question is not completely trivial because, when
$\O\not=0$, the energy is not conserved. Nevertheless, when $\a=0$
the unreduced system has the first integral
$$
\begin{aligned} 
  E(x,v,\o_\z)
  =&
  \frac12|v|^2
  + \frac12\Big(\!\PS xv \psi'\big({\textstyle\frac12}|x|^2\big) \Big)^2
  + \frac k2 \o_\z^2 
  -\O (x_1v_2-x_2v_1) +k\O \o_\z
  \\
  &+\frac k2 \Big((v_1+\O x_2)F(|x|)
   + x_2(\O-\o_\z)\psi'({\textstyle\frac12}|x|^2)\Big)^2
  \\
  &+\frac k2 \Big((v_2-\O x_1)F(|x|)
   - x_1(\O-\o_\z)\psi'({\textstyle\frac12}|x|^2) \Big)^2 
  +  \hat g \psi\big(\textstyle{\frac{|x|^2}2}\big) 
\end{aligned} 
$$
which coincides with the energy for $\O=0$ and, for $\O\not=0$, is
called a `moving energy'.
The existence of this integral for $\O\not=0$ was proven in
\cite{FS2016} and its expression was computed in \cite{BMB2015}. This
function coincides with the function variously called ``energy'',
``total energy'', ``Jacobi integral'' in Lagrangian mechanics but the
fact that---under certain conditions---it is a first integral for
nonholonomic systems with constraints which are affine (linear
nonhomogenous) in the velocities was proven only very recently. We
refer to \cite{FS2016,BMB2015,FGNS2018} for the theory of moving
energies in nonholonomic mechanics.

The impossibility of blow ups is certainly ensured by the compactness of the
level sets of the moving energy, which intuitively prevents $\o_\z$ to
``go to infinity'' and, more precisely, ensures the completeness of the
dynamical vector field. The compactness of {\it all} the level sets of $E$
in the $\SO3\times\SO2$-reduced system was proven in \cite{dVFS},
Proposition 7, in the case $D=\bR2$,
under the hypothesis that the profile function goes
to $+\infty$ at infinity, and does it sufficiently fast if $\O\not=0$.
Due to the compactness of $\SO2$ and $\SO3$, this result extends to
the $\SO3$-reduced system and to the unreduced one. However if, at
infinity, the profile function goes to $-\infty$ or is
bounded, then certainly there are level sets of the moving energy
which reach infinity in the factor $\bR2\ni x$ of $M_8$ and are
not compact. 

Nevertheless, as we show here, there cannot be blow ups at the vertex.
This is due to the fact that, on each level set of $E$, the
coordinates $v$ and $\o_\z$ cannot go to infinity near $x=0$:

\begin{proposition}
\label{p:blowup}
Assume $\a=0$. Then, for any $\O\in\reali$ and any $L>0$ the level
sets of $E$ have  compact intersection with the subset of $M_5$ where
$|x|\le L$.
\end{proposition}

\begin{proof}
Consider $E_0\in\reali$ such that the set $S_{E_0}=\{(x,v,\o_\z)\in
M_5: E(x,v,\o_\z) = E_0\,\; |x|\le L\}$ is not empty. Since $E$ is
continuous, $S_{E_0}$ is closed and we need to prove that it is
bounded. Note that
$$
\begin{aligned}
  E_0 = E(x,v,\o_\z)
  &\;\ge\;
   \frac12|v|^2 +\frac k2 \o_\z^2
   - |\O|\,|x|\,|v|    -k |\O|\,|\o_\z| 
   -\hat g\psi\big(\textstyle{\frac12 |x|^2}\big) 
  \\
  &\;=\;
   \frac12\big(|v|-|\O|\,|x|\big)^2
   + \frac k2 \big(|\o_\z|-|\O|\big)^2 
   - \frac12\O^2 |x|^2 -\frac12 k \O^2 
   -\hat g\psi\big(\textstyle{\frac{|x|^2}2}\big) 
  \,.
\end{aligned}
$$
Hence, for $|x|\le L$,
$$
  E_0
  \ge
   \frac12\big(|v|-|\O|\,|x|\big)^2
   + \frac k2 \big(|\o_\z|-|\O|\big)^2 
   - C
$$
with $C=\frac12(k+L^2)\O^2+\max_{0\le r\le L}|f(r)|$. Thus,
$(|v|-|\O|\,|x|\big)^2 + k \big(|\o_\z|-|\O|\big)^2 \big)\le
2(E_0+C)$ and so $|v|\le L|\O|+\sqrt{2(E_0+C)}$ and
$|\o_\z|\le |\O|+\sqrt{\frac2k(E_0+C)}$.
\end{proof}

\subsection{Linearization at the vertex. }

We study now the possibility that motions tend asymptotically to the
vertex. To simplify the exposition we say that an eigenvalue of the
linearization is of type $Z$ if it is zero, of type $C$ if it is
purely imaginary and nonzero, of type $R_+$ ($R_-$) if it is real and
positive (negative) and of type $F_+$ ($F_-$) if it has nonzero
imaginary part and positive (negative) real part. 

As is well known, the presence of only eigenvalues with zero real
part, hence of types $Z$ and $C$, is a necessary condition for
Lyapunov stability called ``spectral stability''. The presence of
some eigenvalue with positive real part, namely of types $R_+$ and
$F_+$, implies Lyapunov instability. 

But foremost, we are interested in the existence of motions which are
asymptotic, in the future or in the past, to the equilibria at the
vertex, which are related to the presence of eigenvalues of types
$R_-$, $F_-$ and $R_+$, $F_+$, respectively.

We may limit our analysis to the $4\times 4$ block
\eqref{eq:linearisation_5} of the linearization. Obviously, its
complex eigenvalues come in conjugate pairs, but further limitations
come from the biquadratic structure of the characteristic polynomial
\eqref{cp}.

\begin{proposition}
\label{p:LinVertex}
Assume $\a=0$ and define the function
$$
  B(\o_\z,\O) = \big(1+f''(0)\big)\O -f''(0)\o_\z \,.
$$
Then, for any $\O\in\reali$, the four eigenvalues of
the $4\times4$ block \eqref{LinVertex} of the linearization
at the reduced equilibrium $(0,0,\o_\z)$ are of the following types:
\bList
\item[i.] If $f''(0)=0$: $ZZZZ$ if $\O=0$ and $ZZCC$ if $\O\not=0$.
\item[ii.] If $f''(0)>0$: $CCCC$.
\item[iii.] If $f''(0)<0$: $CCCC$ if $B(\o_\z,\O)^2\ge
4\gamma\mu^{-2}|f''(0)|$,
$F_+F_+F_-F_-$ if $0<B(\o_\z,\O)^2<4\gamma\mu^{-2}|f''(0)|$,
and $R_+R_+R_-R_-$ if $B(\o_\z,\O)=0$.
\eList
\end{proposition}

\begin{proof}
Preliminarily note that, if $c\ge0$, then the four roots of the
biquadratic equations $\lambda^4 + 2b\lambda^2 + c=0$ are of the
following types. If $c=0$: $ZZZZ$ if $b=0$, $ZZCC$ if $b>0$,
$ZZR_+R_-$ if $b<0$. If $c>0$: $F_+F_+F_-F_-$ if $b^2<c$,
$R_+R_+R_-R_-$ if $b^2\ge c$ and $b<0$, $CCCC$ if $b^2\ge c$ and
$b>0$. 

When $\a=0$, the four coefficients \eqref{a} evaluated at the
equilibrium $(0,0,\o_\z)$ are $a_{31}=a_{42}=-\gamma f''(0)$ and
$a_{34}=-a_{43}=-\mu B(\o_\z,\O)$ (use $\psi'(0)=f''(0)$, $F(0)=1$).
Therefore, the characteristic polynomial
\eqref{cp} is $\lambda^4+2b\lambda^2+c$ with
$$
   b=\gamma f''(0)+\frac12\mu^2B(\o_\z,\O)^2 \,,\qquad
   c=\big(\gamma f''(0)\big)^2 \,.
$$

(i.) If $f''(0)=0$ then 
$B(\o_\z,\O)=\O$ and so $b=\frac12(\mu\O)^2$ and $c=0$. If $\O=0$
then $b=0$ and the eigenvalues type is $ZZZZ$. If $\O\not=0$ then
$b>0$ and the eigenvalues type is $ZZCC$.

(ii.) If $f''(0)>0$ then $c>0$ and, since $B(\o_\z,\O)^2\ge0$, $b\ge
\gamma f''(0)>0$ and $b^2\ge \big(\gamma f''(0)\big)^2=c$. Thus, the eigenvalues type is $CCCC$.

(iii.) Assume $f''(0)<0$ and write $B$ for $B(\o_\z,\O)$. Thus 
$b=\frac12\mu^2B^2-\gamma|f''(0)|$, $c=(\gamma|f''(0)|)^2>0$ and 
$$
   b^2-c
   =
   \big(b+\gamma|f''(0)| \big) \big(b-\gamma|f''(0)|\big)
   =
   \frac14\mu^2B^2 \big(\mu^2B^2-4\gamma|f''(0)|\big) \,.
$$
We now distinguish two cases. (1) If $\mu^2B^2-4\gamma|f''(0)| \ge0$ then
$b^2-c\ge0$ and, since $b=\frac12\mu^2B^2-\gamma|f''(0)|\ge\gamma|f''(0)|>0$,
the eigenvalues
type is $CCCC$. (2) If $\mu^2B^2-4\gamma|f''(0)|<0$ and $B\not=0$ then
$b^2-c<0$ and the eigenvalues type is $F_+F_+F_-F_-$. If instead $B=0$
then $b^2-c=0$, $b=-\gamma|f''(0)|<0$ and the eigenvalues type is
$R_+R_+R_-R_-$.
\end{proof}

Proposition \ref{p:LinVertex} implies that when $f''(0)\ge0$
all reduced equilibria at the vertex are spectrally stable.

Instead, when $f''(0)<0$, namely the surface has a nondegenerate
maximum at the vertex, the situation is richer. In such a case
$B(\o_\z,\O)= (1-|f''(0)|)\O + |f''(0)|\o_\z$, with $1-|f''(0)|>0$
because of \eqref{eq:curvature}, the loci
$B(\o_\z,\O)=\mathrm{const}$ in the $(\o_\z,\O)$-plane are straight
lines, and the regions of different eigenvalues types are as in
Fig. 2. Therefore, for fixed $\O$, the spectrally stable reduced
equilibria $(0,0,\o_\z)$ are those with $\o_\z$ outside an open
bounded interval (which depends on $\O$, and may include $\o_\z=0$).
In particular, when $\O=0$, the spectrally stable reduced equilibria
are those with $|\o_\z|\ge\frac2\mu\sqrt{\frac\gamma{|f''(0)|}}$.
Interestingly, each reduced equilibrium $(0,0,\o_\z)$ becomes
eventually spectrally stable for $|\O|$ large enough. In this sense,
the rotation of the surface has a ``stabilizing'' effect---a
phenomenon of which some instances had already been pointed out in
\cite{dVFS}.


\begin{figure}[h]
\begin{center}
{\small
{\scalebox{.7}{\includegraphics*{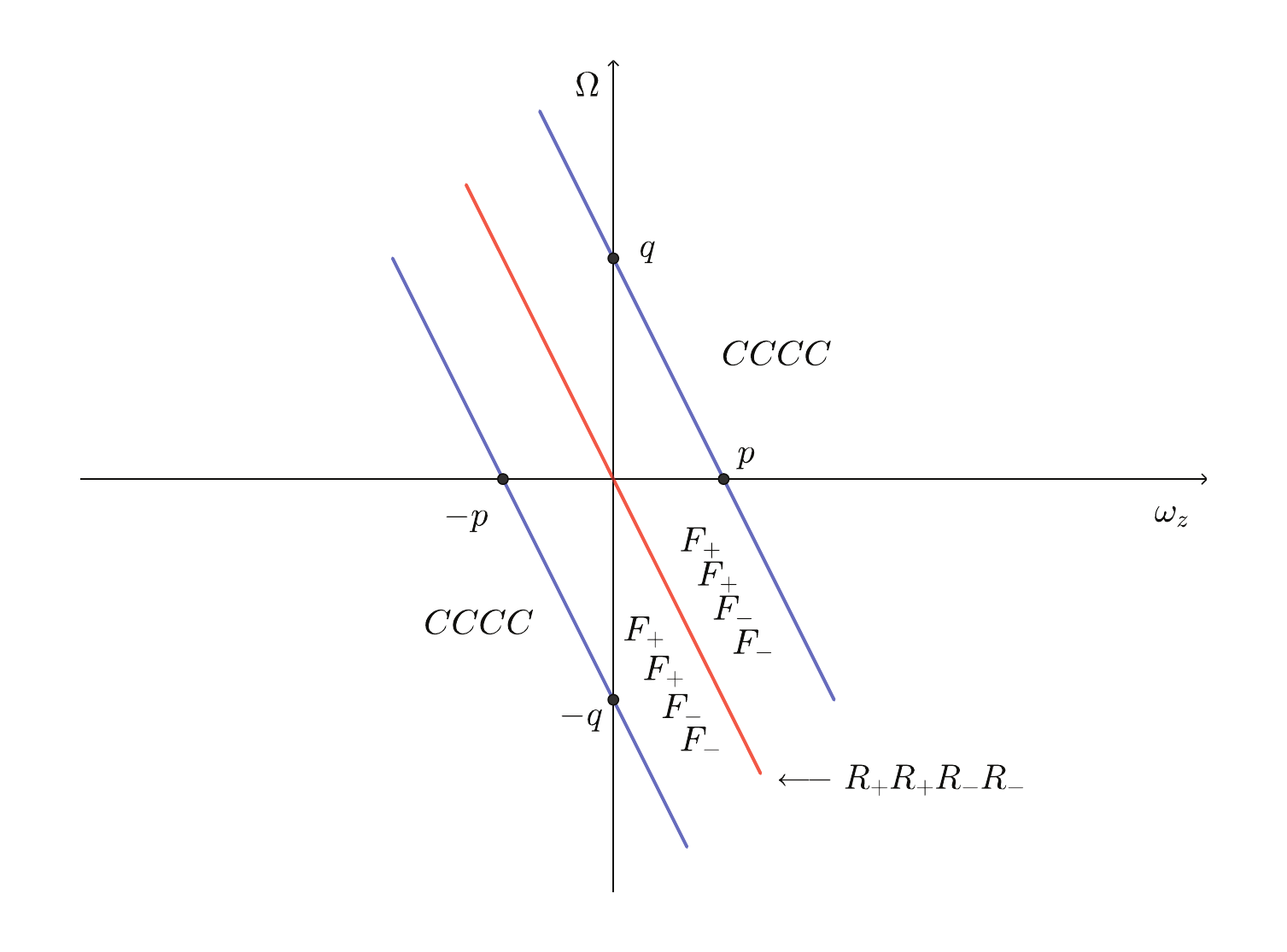}}}
}
\caption{\small The types of the four eigenvalues of the block
\eqref{LinVertex} when $f''(0)<0$ as functions of $\o_\z$ and $\O$. The
marked points are $p=\Big(\frac2\mu\sqrt{\frac\gamma{\,|f''(0)|}},0\Big)$ and
$q=\Big(0,\frac{2\sqrt{\gamma |f''(0)|}}{\mu(1-|f''(0)|)}\Big)$.}
\end{center}
\end{figure}




But moreover, when $f''(0)<0$, for $(\o_\z,\O)$ in the instability
region
\begin{equation}
\label{InstReg}
  -2\mu^{-1} \sqrt{\gamma|f''(0)|}
  < \big(1+f''(0)\big)\O -f''(0)\o_\z 
  < 2\mu^{-1} \sqrt{\gamma|f''(0)|} 
\end{equation}
the reduced equilibrium $(0,0,\o_\z)$ at the vertex has a
two-dimensional stable manifold and a two-dimensional unstable
manifold on which all motions tend to the equilibrium for,
respectively, $t\to+\infty$ and $t\to-\infty$. (The existence of these
invariant manifolds is often stated for hyperbolic equilibria, but it
is granted also in the present case because the eigenvalues with
negative (positive) real parts are separated by a ``spectral gap''
from all the others, including zero; see section 4.1 of
\cite{chicone}). Thus, in all motions in these submanifolds, for
either $t\to+\infty$ or $t\to-\infty$ the center of the ball tends
asymptotically to the vertex, with the $\z$-component of the angular
velocity of the ball approaching a finite value. Note that, in 
region (\eqref{InstReg}), the eigenvalues of the $4\times4$ block
(\eqref{eq:linearisation_5}) of the linearization have generically
nonzero imaginary parts. Therefore, in that region, generically
motions will tend to the vertex with some kind of spiraling. Motions
that tend to the equilibrium without spiraling are exceptional
($B(\o_\z,\O)=0$).

\subsection{Lyapunov stability. }

Going beyond the linearized analysis, it would be interesting to study
the Lyapunov stability of the spectrally stable reduced equilibria at
the vertex. The natural candidate for a Lyapunov function is the
moving energy. However, $dE(0,0,\o_\z) = (0,0,k(\o_\z-\O))$ and the
moving energy has a critical point only at those reduced equilibria
$(0,0,\o_\z)$ with $\o_\z=\O$ (the ball stands still relative to the
rotating surface, but spins in space). We restricts our
considerations to this case.

\begin{proposition}
\label{p:vertex} Assume $\a=0$ and $\O\in\reali$. If $f''(0)>0$ and
$\O^2<\hat g f''(0)$ then the reduced
equilibrium $(0,0,\o_\z=\O)$ is Lyapunov stable. 
\end{proposition}

\begin{proof}
Lyapunov stability of $(0,0,\o_\z=\O)$ is granted if the
Hessian
\begin{equation}
\label{LinVertex}
  \begin{pmatrix}
     k\O^2+\hat g f''(0) & 0 & 0 & -(1+k)\O & 0 \\
     0 & k\O^2+\hat g f''(0) & (1+k)\O & 0 & 0  \\
     0 & (1+k)\O & 1+k& 0 & 0            \\
     -(1+k)\O & 0 & 0 & 1+k &0            \\
     0 & 0 & 0 & 0 & k 
  \end{pmatrix}
\end{equation}
of the moving energy $E$ at that point is positive definite.
Clearly, its last three principal minors are all positive. The first
two minors equal $k(1+k)(\hat g f''(0)-\o_\z^2)^2$ and $k(1+k)(\hat g
f''(0)-\o_\z^2)$, respectively, and are both positive if $\hat
gf''(0)>\o_\z^2$. 
\end{proof}

This result is somewhat poor, because it applies only to cases in
which the vertex is a point of nondegenerate minimum of the surface,
and only to the equilibria with $\o_z=\O$. It does not allow to say
anything about Lyapunov stability in all other cases. But also
in the considered case, it detects Lyapunov stability only for
$|\o_z|=|\O|$ not too large ($<\!\sqrt{\hat gf''(0)}\,$), while 
in that situation there is spectral stability for all
$\o_z=\O\in\reali$: it would be interesting to establish if
Lyapunov stability of this class of equilibria is retained for all $|\O|$
or if it is actually lost at large $|\O|$ (a sort of gyrostatic
de-stabilization?). Perhaps, a study of Lyapunov stability beyond the
result of Proposition \ref{p:vertex} could be based on trying to
build a Lyapunov function out of the moving energy and of the two
``Routhian'' integrals.

\section{The Kasamawashi case ($\mathbf{\a\not=0}$, $\mathbf{f''\le0}$)}
\label{s:kasam}
We consider now the case in which the surface $\Sigma$ is inclined of
an angle $\a$, $0<\a<\frac\pi2$. Imagining a ball that rolls on
the surface of an umbrella we assume that $f$ is concave, $f''(r)\le
0$ for all $r$. Thus, $f'(r)\le 0$ for all $r>0$ as well. In such a
situation, an equilibrium $(x_1,0,0,0,\o_\z)$ has necessarily $x_1>0$
and $f'(x_1)=-\tan\a<0$. 

\begin{proposition}
\label{p:kasam}
Under the stated hypotheses, let $\cE=(x_1,0,0,0,\o_\z)$
be an equilibrium of the system, with $x_1>0$. 
\bList
\item[i.] If $f''(x_1)=0$, then $\cE$ is spectrally stable if and only
if
\begin{equation}
\label{f''=0}
  \Big(\frac{x_1}{\sin\a}-1\Big)\O^2 + \O \o_\z \ge \frac\gamma{\mu^2}
\,.
\end{equation}
\item[ii.] If $f''(x_1)<0$, define $h:=\frac{f''(x_1)}{f'(x_1)}
=-f''(x_1)\frac{\cos\a}{\sin\a}>0$.
Then, $\cE$ is spectrally stable if and only if 
\begin{equation}
\label{f''<0}
   a_{11}\O^2 + 2a_{12}\O\o_\z + a_{22}\o_\z^2 \ge a_{0}
\end{equation}
with $a_{11}=(x_1-\sin\a)\big(1+h\sin\a-h x_1\sin^2\a\big)$, 
$a_{12}=\frac12\big(1 +2h\sin\a -h x_1-h x_1\sin^2\a\big)\sin\a$,
$a_{22}=-h \sin^2\a$,
$a_{0}=\frac\gamma{\mu^2} \big(1+2\sqrt{h x_1}\cos\a
+ h  x_1\cos^2\a\big)\sin\a$.
\eList
\end{proposition}

\begin{proof}
Since $x_1>0$ and $0<\a<\frac\pi2$, $\sin\a$ and $\cos\a$ are both positive, 
and $f'(x_1)=-\tan\a$. Thus $F(x_1)=\frac1{\cos\a}$, 
$\psi'({\textstyle\frac12}x_1^2)=-\frac{\tan\a}{x_1}$,
$\psi''({\textstyle\frac12}x_1^2)=f''(x_1)-\frac{\tan\a}{x_1^2}$ and
the entries
\eqref{a} of the $4\times4$ block \eqref{eq:linearisation_5} of the
linearization can be written as
\[\begin{aligned} 
   a_{31} = & -\gamma f''(x_1) \cos^3(\a)\,, \qquad   
   a_{34} = - \mu\O + \mu (\O- \o_\z)\frac{\sin\a} {x_1}\,,\\
   a_{42} = & \, \gamma \frac{\sin\a}{x_1} \,, \qquad
   a_{43} = \mu\O + \mu (\O -\o_\z - x_1 \O \sin\a)f''(x_1)\cos\a \,. 
\end{aligned}\]
Spectral stability of $\cE$ is equivalent to the fact that all the roots
of the characteristic polynomial \eqref{cp}, namely
$\lambda^4+2b\lambda^2+c$ with $2b =- (a_{31}+a_{42}+a_{34}a_{43})$
and $c=a_{31}a_{42}$, have nonpositive real part.

(i.) If $f''(0)=0$ then $c=0$ and, as noticed in the proof of
Proposition \eqref{p:vertex}, the roots of the characteristic
polynomial have all nonpositive real part if and only if $b\ge0$. For
$f''(0)=0$, $2b=\mu^2\big(1-\frac{\sin\a}{x_1}\big) \O^2 +
\mu^2 \frac{\sin\a}{x_1}\O\o_\z -\gamma \frac{\sin\a}{x_1}$. Since
$x_1>0$, $\sin\a>0$ and $\mu>0$, condition $b\ge0$ is equivalent to
\eqref{f''=0}.

(ii.) If $f''(0)<0$ then $c>0$ and (see again the proof of
Proposition \eqref{p:vertex}) the roots of the characteristic
polynomial have all nonpositive real part if and only if $b^2\ge c$
and $b>0$, namely $b\ge\sqrt{c}$. Since $x_1>0$ and, as noticed above,
$f'(x_1)=-\tan\a<0$, $h :=\frac{f''(x_1)}{f'(x_1)}>0$. Writing
$f''(x_1)=- h\tan\a$, condition $b\ge\sqrt c$ becomes
$\frac{\mu^2}{2x_1}(a_{11}\O^2+2a_{12}\O\o_\z+a_{22}\o_\z^2-a_{00})\ge0$.
\end{proof}

We now analyze the conditions given by Proposition \ref{p:kasam}. 

Given $x_1$, when $f''(x_1)=0$ the condition of spectral stability
\eqref{f''=0} is satisfied in a region of the $(\o_\z,\O)$-plane which is
bounded by the two branches of a hyperbola and is shown in Fig. 3.
One asymptote of the hyperbola is the $\o_\z$-axis, and the
equilibrium is never spectrally stable (and hence is always unstable)
if $\O=0$. The rotation of the
surface has a stabilizing effect, in the sense that if $\O\not=0$
then spectral stability of the equilibrium becomes possible for
certain $\o_\z$, but this effect depends on the distance of the
equilibrium position from the rotation axis. Indeed, the other
asymptote of the hyperbola is the line
$\O=(1-\frac{\sin\a}{x_1})\o_\z$ and counterclockwise rotates from the
diagonal to the horizontal axis as $x_1$ grows from $0$ to $+\infty$.

Thus, for equilibria near the rotation axis ($x_1<\sin\a$) spectral
stability is achieved for $\o_\z$ of the same sign as $\O$ and in an
unbounded interval which does not contain $0$, and whose size first
decreases and then increases with $|\O|$. 

Instead, for equilibria far from the rotation axis ($x_1>\sin\a$),
spectral stability is achieved for $\o_\z$ in an interval that
contains $0$ and whose size steadily increases as $|\O|$ increases. 

As already mentioned in the Introduction, the case $f''(x_1)=0$ is
that of the kasamawashi, which uses an umbrella with conic profile.
The umbrella is inclined so that the upper generatrix of the
cone is horizontal, and there are reduced equilibria at all points of
this horizontal line. 
Inspection of movies showing actual kasamawashi
performances\footnote{Such as the one available at {\tt
https://www.youtube.com/watch?v=FeDyMdh1JLQ}} suggests that the
performer manages to have $\o_\z=0$ and that, consistently with the
above remarks, $x_1>\sin\a$.\footnote{In the movie, the angle $\a$ is
small and the ball sits at a distance from the rotation axis which is
approximately two-to-three times its radius, hence $x_1>1$.} Of
course, these conclusions should be taken for what they are
because---besides the fact that, as already pointed out, kasamawashi
involves control---not only spectral stability does non guarantees
stability but, moreover, the presence of zero
eigenvalues might be an indication of unstable behaviours. Some further
study of the dynamics might be interesting.


\begin{figure}[h]
\begin{center}
{\small
{\scalebox{.5}{\includegraphics*{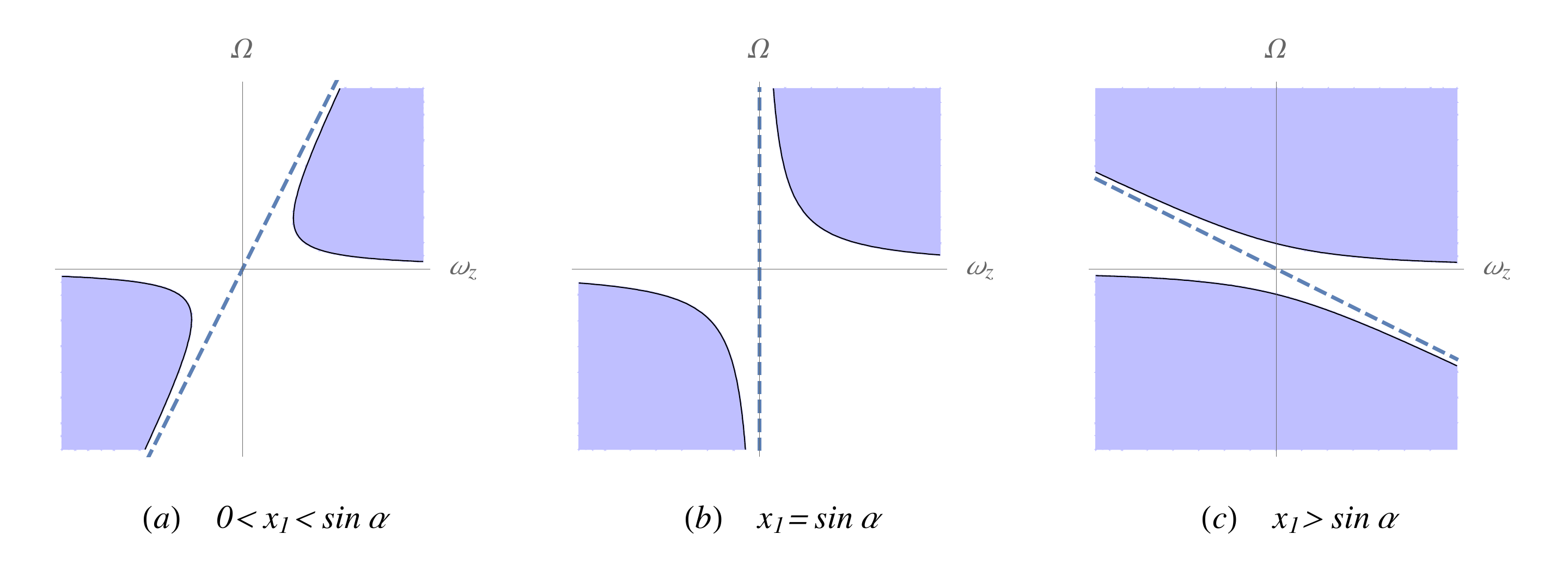}}}
}
\caption{\small The region of spectral stability of the equilibrium
$(x_1,0,0,\o_\z)$ in the plane $(\o_\z,\O)$  when $f''(x_1)=0$. The dashed line is
the asymptote $\O=\frac{\sin\a}{\sin\a-x_1}\o_\z$}
\end{center}
\end{figure}




When $f''(x_1)<0$ the situation is similar, though more complex to
analyze. First, when $\O=0$ condition \eqref{f''<0} reduces
to 
\begin{equation}
\label{f''<0,O=0}
 \o_\z^2 \ge 
 \frac\gamma{\mu^2}
 \Big(\frac1h +\sqrt{\frac{x_1}h }\cos\a + x_1\cos^2\a\Big) \,.
\end{equation}
Therefore, at variance from the case $f''(x_1)=0$, for $\O=0$ there is
spectral stability for $|\o_\z|$ not too small (with a threshold which
however increases with $x_1$). For all $\O$,
$$
  a_{11}a_{22}-a_{12}^2 = 
  -\frac14\mu^2\big(2+h x_1\cos(2\a)\big)^2\sin^2\a 
$$
is negative (unless $h x_1\cos(2\a)=-2$, which could only happen if
$\a\ge\frac\pi4$) and region \eqref{f''<0} is again bounded by the two
branches of a hyperbola. These curves intersect the $\o_\z$-axis in the
two points where \eqref{f''<0,O=0} is satisfied with the $=$ sign.
From this it follows that the region where \eqref{f''<0} is satisfied
is the one outside the two branches of the hyperbola---very much as in
Fig. 3.

\begin{remark}
If $f''(r)<0$ for all $r$ then for any
$\a\in(0,\frac\pi2)$ there is a unique $\o_\z$-family of equilibria
$(x_1,0,0,0,\o_\z)$. For $\a\to0$, these equilibria tend to the equilibria
$(0,0,0,0,\o_\z)$ at the vertex. It is not difficult to check that, for
small $\a$, at first order in $\a$ the condition for spectral stability 
\eqref{f''<0} coincides with the condition $B(\o_\z,\O)\ge 4\gamma|f''(0)|$
which, in item iii. of Proposition \ref{p:LinVertex}, ensures the
spectral stability of the equilibria at the vertex. (Since 
$\sin\a\sim\a$ etc, $f'(x_1)\sim\a$ and $f'(x_1)\sim f''(0)x_1$ which
give $x_1\sim \frac\a{|f''(x_1)|}$ and $h\sim\frac1\a$).  
\end{remark}

\section{Conclusions } We have studied two new problems in the
dynamics of a heavy homogeneous ball that rolls without sliding on a
surface of revolution which rotates with constant angular velocity
$\O\in\reali$ about its figure axis. The system has an
$\SO3$-invariance which allows reduction to 5-dimensions. 

First, assuming that the figure axis of the surface is vertical, we
have studied those equilibria of the reduced system which correspond
to periodic orbits of the unreduced system in which the ball sits at
the vertex of the surface and rotates steadily about its center with
vertical angular velocity $\o_\z\in\reali$. We have shown that no
blow up is possible at these reduced equilibria and we have studied
their spectral stability as a function of the parameters, in
particular of $\o_z$, $\O$ and the curvature of the surface's profile
at the vertex. We have shown that they are all spectrally
stable unless the profile of the surface has a nondegenerate maximum
at the vertex, in which case spectral stability is attained for
$(\o_z,\O)$ outside of a strip in $\bR2$. For $(\o_z,\O)$ inside
that strip the reduced equilibrium is spectrally unstable, and
this implies the existence of motions which are asymptotic (in the
past or in the future) to the reduced equilibrium. Finally, we have
proven the nonlinear stability of a special subclass of the
spectrally stable reduced equilibria: in the case in which the
surface has a nondegenerate minimum at the vertex, those with
$\o_z=\O$ and $|\O|$ not too large. It is likely that the class of
nonlinearly stable reduced equilibria at the vertex is larger, but
this question remains open and deserves to be studied.

Second, we have considered the case in which the figure axis is
tilted with respect to the vertical. The reduced equilibria correspond
to periodic motions of the unreduced system in which the ball
steadily rotates with vertical angular velocity $\o_z$ about its
center, which stands still in space over a point in which the surface
has horizontal tangent plane. We have limited the  study of the spectral
stability of these reduced equilibria to the case of a non-convex
profile, a particular case of which is that of the conic umbrella
used in the kasamawashi performances, remarking in particular its
dependence on the distance from the vertex. A study of the nonlinear
stability of these reduced equilibria, and even more so of the
dynamics near them, is left open and is worth further investigation.

\section*{Appendix: the equations of motion of the system}

\setcounter{equation}{0}
\renewcommand{\theequation}{{\rm A}.\arabic{equation}}

The equations of motion of the system can be determined in various
routine ways which however, as often happens with nonholonomic
systems, involve some tedious computations. Here we follow the
approach of \cite{dVFS}.

Reference \cite{dVFS} employs a known form of the equations of motion of
mechanical nonholonomic systems as the restriction to the constraint
manifold of Lagrange equations with the nonholonomic reaction forces,
writing however them in a way that allows for the use of
quasi-velocities (Proposition 16 in the Appendix of \cite{dVFS}). Of
course, one might just specialize those formula to the present case,
and this would indeed be the straightest---though somewhat
laborious---approach. However, since the
computations are there already made for the case $\a=0$, in order to
keep the length of this article to a minimum we prefer here to indicate how to modify
that deduction to allow for $\a\not=0$. There are in fact three other
minor differences. One is technically irrelevant: reference
\cite{dVFS} assumes that the domain $I=(-R,R)$ of the profile function is
the entire real axis, so that $D=\bR2$. In addition, the
derivation of the equations of motion in the Appendix of \cite{dVFS}
uses the profile function $f$, not $\psi$, and a different
parametrization of $M_8$, which excludes the vertex and uses polar
coordinates, namely
$(r,\theta,v_r,v_\theta,\cR,\o_\z)\in\reali_+\times S^1\times
\bR{}\times \bR{}\times\SO3\times\bR{}=:M_8^\mathrm{pol} $ with
$x_1=r\cos\theta$, $x_2=r\sin\theta$, $v_r=\dot r$,
$v_\theta=\dot\theta$. We thus indicate how to modify such a
derivation.

First, the inclination of the surface has the only effect of changing
the potential energy of the weight force: instead of $g\z|_{M_8^\mathrm{pol}}
=a \hat gf(r)$, it becomes $g(\z\cos\a+\x\sin\a)|_{M_8^\mathrm{pol}} =
a\hat g (f(r)\cos\a + r\sin\a \cos\theta )$. This has the consequence
that the nonholonomic reaction force $R$, given in formula (46)
within the proof of Proposition 17 of \cite{dVFS}, gets the
following changes: in its $\dot r$-component the term $\mu\hat g f'$
has to be replaced with $\mu \hat g (f' \cos\a + \sin\a \,
\cos\theta)$, its $\dot\theta$-component acquires a term $-\mu\hat g
r^{-1}\sin\a\,\sin\theta$ and its $\o_\z$-component acquires a term
$-\mu\hat g F^{-1}f'\sin\a\,\sin\theta$. These changes propagate to
the equations for $\dot v_r$, $\dot v_\theta$ and $\dot\o_\z$ as
given in Proposition 17 of \cite{dVFS} after
multiplication by the appropriate entries of the inverse of the
kinetic matrix (namely $F^{-2}$, $r^{-2}$ and $k^{-1}$ respectively).

Second, the equations for $\dot v_r$ and $\dot v_\theta$ can be
transformed into equations for $\dot v_1$ and $\dot v_2$ using the
kinematical identities $\dot v_1 = \big(\frac{\dot
v_r}r-v_\theta^2\big)x_1 -\big(\dot v_\theta + 2\frac{v_r
v_\theta}r\big)x_2$ and $\dot v_2 = \big(\frac{\dot
v_r}r-v_\theta^2\big)x_2 + \big(\dot v_\theta +
2\frac{v_rv_\theta}r\big)x_1$ and making the obvious substitutions
$r\to|x|$, $\sin\theta\to\frac{x_2}r$, $\cos\theta\to\frac{x_1}r$,
$v_r\to x\cdot v$, $v_\theta\to\frac{x_1v_2-x_2v_1}r$. This leads to
the equations $\dot x_1=v_1$, $\dot x_2=v_2$, $\dot\cR =\cR^T\o$ and
$$
\begin{aligned} 
  \dot v_1 =
  &  
  - \frac\gamma{F^2} \bigg( \frac{x_1}{|x|}f'\cos\a +
     \Big(1 + \frac{x_2^2}{|x|^2} f'^2 \Big)\sin\a\bigg)
  + \frac\mu F \Big( \frac{x_1}{|x|^3} \PS x{Jv} f'   
             + \frac{x_2}{|x|^2} \PS xv f'' \Big)\o_z
  \\
  &
  - \frac\mu{F^2} \frac{v_1}{|x|} \PS xv f' f''
  - \frac{f'}{(1+k)F^2} \frac{x_1}{|x|^4} 
      \Big( (\!\PS x{Jv}\!)^2f'+ |x|(\!\PS xv\!)^2f''\Big) 
   \\
  &
  - \O \mu \bigg(v_2 + \frac1F \frac{x_1}{|x|^3} \PS x{Jv} f' 
     + \frac{x_2}{|x|^2}\frac{\PS xv}{F^2} f''
        \big(F + |x|f'\big)\bigg) 
  \\
  \dot v_2 =
  &  
  - \frac\gamma{F^2} \frac{x_2}{|x|} f'
        \Big(\cos\a - \frac{x_1}{|x|}f'\sin\a\Big)
  + \frac\mu F \Big(\frac{x_2}{|x|^3}  \PS x{Jv} f'
             - \frac{x_1}{|x|^2} \PS xv f''\Big) \o_z
  \\
  &
  - \frac\mu {F^2} \frac{v_2}{|x|} \PS xv f' f'' 
  - \frac{f'}{(1+k)F^2}\frac{x_2}{|x|^4}
       \Big( (\!\PS x{Jv})^2 f' + |x|(\!\PS xv)^2f''\Big) 
  \\
  &
  + \O \mu \bigg( v_1- \frac1F\frac{x_2}{|x|^3}\PS x{Jv} f' 
    + \frac{x_1}{|x|^2} \frac{\PS xv}{F^2}f''\big(F+|x|f'\big)
  \bigg)
  \\
  \dot \o_\z =
  & 
   -\frac\gamma F \frac{x_2}{|x|}f'\sin\a
   - \frac{f'f''}{(1+k)F^3} \frac{\PS xv}{|x|^2}
     \big(|x|F\o_\z- \PS x{Jv} f' \big)
  \\
  &
   + \O \frac{f'}{(1+k)F} \frac{\PS xv}{|x|} 
     \Big(1+ \frac{f''}F +|x|\frac{f'f''}{F^2} \Big)
\end{aligned}
$$
with $J=\left( \begin{matrix} 0&1 \\ -1&0 \end{matrix}\right)$.
After replacing $f'$ with $|x|\psi'$ and $f''$ with 
$\psi'+|x|^2\psi''$, see \eqref{psi'psi''}, these 
equations take the form \eqref{eq:eqmoto}. 

In this way we have proven that \eqref{eq:eqmoto} are the equations
of motion of the system in the subset of the phase space $M_8$ where
$x\not=0$. Therefore, their right hand side defines a vector field
$Y$ in $M_8\setminus\{x=0\}$ which coincides with the restriction to
such a set of the dynamical vector field of the system. But since the
latter is known (from the general theory) to exists in all of $M_8$,
$M_8\setminus\{x=0\}$ is dense in $M_8$ and $Y$ has a continuous
extension to $M_8$, the extension of $Y$ is the dynamical vector
field of the system in all of $M_8$.

\vspace{2ex}\noindent
{\bf Acknowledgements.} We would like to thank Prof. Toshiro Iwai for
pointing out to one of us the similarity between the dynamics of the
ball on a rotating surface and the kasamawashi art. FF has been
partially supported by the MIUR-PRIN project 20178CJA2B {\it New
Frontiers of Celestial Mechanics: theory and applications}.

\end{document}